\def\b{\bibitem}
\def\be{\begin{equation}}
\def\ee{\end{equation}}
\def\bea{\begin{eqnarray}}
\def\eea{\end{eqnarray}}
\def\bml{\begin{mathletters}}
\def\eml{\end{mathletters}}
\documentstyle[aps,prl,epsf,floats]{revtex}

\begin{document}
% Macros for the various macro package names, etc.
\def\SNG{{\em Physical Review Style and Notation Guide}}
\def\LUG {{\em \LaTeX{} User's Guide \& Reference Manual}}
\def\btt#1{{\tt$\backslash$\string#1}}%
\def\REVTeX{REV\TeX}
\def\AmS{{\protect\the\textfont2
        A\kern-.1667em\lower.5ex\hbox{M}\kern-.125emS}}
\def\AmSLaTeX{\AmS-\LaTeX}
\def\BibTeX{\rm B{\sc ib}\TeX}
%\makeatletter
%\tighten
\twocolumn[\hsize\textwidth\columnwidth\hsize\csname@twocolumnfalse%
\endcsname
\title{Transport Anomalies and Marginal Fermi-Liquid Effects\\
               at a Quantum Critical Point\\
       \small{$[$ Phys. Rev. Lett. {\bf 85}, 4602 (2000) $]$}}
\author{D. Belitz$^1$, T.R. Kirkpatrick$^2$, R. Narayanan$^3$, and
        Thomas Vojta$^4$}
\address{$^1$Department of Physics and Materials Science Institute,
         University of Oregon, Eugene, OR 97403}
\address{$^2$ Institute for Physical Science and Technology and Department
         of Physics,\\ University of Maryland, College Park, MD 20742}
\address{$^3$ Theoretical Physics, University of Oxford, OX3 1NP, UK}
\address{$^4$ Institut f{\"u}r Physik, TU Chemnitz, D-09107 Chemnitz, Germany}
\date{\today}
\maketitle
\begin{abstract}
The conductivity and the tunneling density of states of disordered itinerant 
electrons in the vicinity of a ferromagnetic transition at low temperature are 
discussed. Critical fluctuations lead to nonanalytic frequency and temperature 
dependences that are distinct from the usual long-time tail effects in a 
disordered Fermi liquid. The crossover between these two types of behavior is 
proposed as an experimental check of recent theories of the quantum 
ferromagnetic critical behavior. In addition, the quasiparticle properties at 
criticality are shown to be those of a marginal Fermi liquid.
%
%% 591 characters
%
\end{abstract}
]
%\narrowtext
Long-range spatial and temporal correlations are characteristic of
systems near critical points, where 
critical soft modes lead to power-law, or scale invariant, behavior 
of correlation functions\cite{Fisher}. This holds primarily for the 
order parameter correlations, but correlation functions of other 
observables that couple to the order parameter show related effects
at a critical point. Away from critical points, soft modes unrelated 
to a phase transition can induce scale invariance in entire
regions of parameter space. Such `generic scale invariance' is in 
general stronger in quantum systems than in classical ones\cite{us_ernst}.

An example of generic scale invariance in quantum systems are the
`weak-localization' effects in disordered metals, which consist of nonanalytic 
temperature, frequency, and wavenumber dependences of various 
observables\cite{LeeRama}. For
instance, in three-dimensions ($3$-$D$) the density of states (DOS) 
as a function 
of bias voltage has a square-root singularity at the Fermi level, 
the conductivity varies like the square root of temperature,
etc. These nonanalyticities provide corrections to the leading disordered 
Fermi-liquid behavior in the metal. They arise from the diffusive nature
of the basic fermionic excitations which couple to the various observables.
In perturbation theory, integrals over diffusion poles lead to 
nonanalyticities\cite{LeeRama}. Alternatively, these phenomena can 
be considered as
corrections to scaling near a disordered Fermi-liquid fixed 
point\cite{us_fermions}. In $2$-$D$, the corresponding effects
are stronger and destroy the Fermi liquid. Near a quantum critical
point, critical fluctuations appear. In general, they have a different
dispersion relation then diffusive modes, and the nonanalyticities induced
by them can
be strong enough to destroy the Fermi-liquid behavior even in $3$-$D$.
Non-Fermi liquid behavior induced by quantum critical points
has been of much interest lately, in particular in connection with an
antiferromagnetic quantum critical point that has been proposed to underly
some of the behavior of high-T$_c$ superconductors\cite{non-FL}.

In this Letter we study the effects of a {\em ferromagnetic} (FM) quantum
critical point on the transport, tunneling, and quasiparticle (QP)
properties of disordered metals. The thermodynamic properties at this phase
transition have been discussed elsewhere\cite{us_dirty,us_paper_I,us_paper_II}.
We are motivated by the following observations: (1) The quantum FM
transition is the best studied of all
quantum critical points\cite{Hertz}. Recent theoretical work\cite{us_dirty}
has shown that the `weak-localization' effects mentioned above couple to the
quantum critical behavior and produce long-range interactions between spin
density fluctuations. The most important implication is
that these effects fundamentally modify the mean-field critical behavior 
predicted by Hertz\cite{Hertz}. However, the feedback of the critical
behavior on the fermionic degrees of freedom, and in particular on the
`weak-localization' effects, has been investigated only to the extent that
it is needed to determine the exact critical behavior. For instance,
the behavior of the resistivity across the quantum FM critical
point is not known theoretically. (2) There are examples of itinerant 
ferromagnets, for instance MnSi, whose Curie temperatures under ambient
conditions are very low due to doping or alloying with nonmagnetic
materials. Such materials can be driven through the quantum ($T=0$) 
transition by means of stress tuning\cite{Lonzarich}. The quantum critical 
point is thus readily accessible. While these experiments could in principle
determine, e.g., the magnetization as a function of the distance from
the critical point, and thus provide a direct check of the theory put
forward in Ref.\ \cite{us_dirty}, in practice this is difficult to do.
On the other hand, transport measurements can be easily
performed\cite{Lonzarich}.

Given points (1) and (2) above, our motivation for this Letter is to
report theoretical results on observables that qualitatively depend on
the quantum critical behavior, but are more easily accessible experimentally
than a direct determination of the latter. Such observables
include the resistivity, and the tunneling DOS. We will also
discuss the dephasing time, which provides a finite-temperature cutoff for 
various singularities in disordered electron systems, and the QP inelastic 
lifetime, which is of fundamental theoretical interest\cite{CKL}.

We start by stating our results. We will consider the following
observables: The electrical conductivity $\sigma$, the tunneling DOS
$N$, the phase relaxation time $\tau_{\rm ph}$, and the
QP lifetime $\tau_{\rm QP}$. Let $t$ be the dimensionless distance
from the quantum critical point at zero temperature, $T=0$,
$\Omega$ the frequency, and
$\epsilon$ the distance in energy space from the Fermi surface.
We will use units such that $\hbar = k_{\rm B} = 1$. Let the disorder
strength be characterized by the elastic mean-free path $\ell$ in units of
the inverse Fermi wavenumber $k_{\rm F}$.
In $3$-$D$ we find at criticality, $t=0$,
\bml
\label{eqs:1}
\bea
\sigma(T)&=&\sigma_0\,\left[1 + c_{\sigma}
  \left(\frac{T}{T_{\rm F}}\,g(\ln (\frac{\epsilon_{\rm F}}{T}))\right)^{1/3}
%\nonumber\\
%\hskip 78pt 
\hskip -7pt + O(\sqrt{T}) \right],
\label{eq:1a}\\
N(\epsilon)&=&N_{\rm F}\left[1 + c_N\left(\frac{\epsilon}{\epsilon_{\rm F}}\,
   g(\ln (\frac{\epsilon_{\rm F}}{\epsilon}))\right)^{1/3}
%\nonumber\\
 \hskip -7pt + O(\sqrt{\epsilon})\right],
\label{eq:1b}\\
\tau_{\rm ph}^{-1}(\epsilon) &=& \epsilon_{\rm F}\left[\,
   c_{\tau}^{\rm ph}\,(\epsilon/\epsilon_{\rm F})\,
          g(\ln(\frac{\epsilon_{\rm F}}{\epsilon})) + O(\epsilon)\right]\ ,
\label{eq:1c}\\
\tau_{\rm QP}^{-1}(\epsilon) &=& \epsilon_{\rm F}\left[\,c_{\tau}^{\rm QP}
   \,(\epsilon/\epsilon_{\rm F})\,
   \ln[\ln(\epsilon/\epsilon_{\rm F})]/\ln(\epsilon/\epsilon_{\rm F})\right.
\nonumber\\
   &&\hskip 28pt + \left. O(\epsilon \ln\ln\ln\epsilon/\ln\epsilon)\right]
       \quad.
\label{eq:1d}
\eea
\eml%
Here $\sigma_0$ and $N_{\rm F}$ are the disordered Fermi-liquid
values of $\sigma$ and $N$, respectively, $\epsilon_{\rm F}$ is the
Fermi energy, and
\bml
\label{eqs:2}
\be
g(x) = \sum_{n=0}^{\infty} [(c(3)\,x)^n/n!]\,e^{(n^2-n)\ln(2/3)/2}\quad,
\label{eq:2a}
\ee
with $c(3) = O(1)$, provides logarithmic corrections to power-law scaling. 
In an asymptotic expansion for large $x$, the leading term is
\be
g(x) \approx \left[2\ln (3/2)/\pi\right]^{-1/2}\,e^{[\ln (c(3)\,x)]^2/
       2\ln (3/2)}\quad,
\label{eq:2b}
\ee
\eml%
The $c_{\sigma}$, $c_N$, and $c_{\tau}$ 
depend on the disorder. For weak disorder, $k_{\rm F}\ell >> 1$,
their leading terms all are of the form $c_i = {\tilde c}_i/k_{\rm F}\ell$, 
with the ${\tilde c}_i$ constants of $O(1)$. 
Equation (\ref{eq:1a}) also holds if $T$ 
is replaced by $\Omega$, and Eqs.\ (\ref{eq:1b}), (\ref{eq:1c}), (\ref{eq:1d}) 
also hold if $\epsilon$ is replaced by $T$. 

\bml
\label{eqs:3}
The terms shown are the exact leading temperature and
energy dependences, within the framework of a perturbative renormalization
group (RG), of these observables at criticality.
At zero temperature, $T=0$, and zero frequency or energy, $\epsilon=0$, we find
\bea
\sigma(t) &=& \sigma_0\left[1 + d_{\sigma}\,t\,g(\ln(1/t))
                  \ln(1/t) + O(t)\right]\quad,
\label{eq:3a}\\
N(t) &=& N_{\rm F}\left[1 + d_N\,t\,g(\ln(1/t))\ln(1/t) + O(t)\right]\quad.
\label{eq:3b}
\eea
For weak disorder, $d_i = O(1)/k_{\rm F}\ell$. The scattering rates
$1/\tau$, of course, vanish at zero energy and temperature. For sufficiently
small values of $\epsilon$ at nonzero $t$ their behavior is given by
\be
\tau_{\rm ph}^{-1}(t,\epsilon)\propto\tau_{\rm QP}^{-1}(t,\epsilon)\propto
(\epsilon/t)^{3/2} \quad.
\label{eq:3c}
\ee
\eml%
These results can be summarized by means of the following generalized 
homogeneity laws, valid for all $2<D<4$,
\bml
\label{eqs:4}
\bea
\sigma(t,T,\Omega)&=&{\rm const.}\times t\,g(\ln(1/t))\,\ln b 
\nonumber\\
&&\hskip -50 pt + F_{\sigma}(tb^{-2}f(b),\Omega f(b),Tf(b), u b^{-(D-2)})\quad,
\label{eq:4a}
\eea
\bea
\Delta N(t,\epsilon,T) &=& {\rm const.}\times t\,g(\ln(1/t))\,\ln b
\nonumber\\
&&\hskip -50 pt + b^{-(D-2)}\,F_N(\epsilon f(b),T f(b)) \quad,
\label{eq:4b}
\eea
\be
\tau_{\rm ph}^{-1}(t,\epsilon,T)\! =\! b^{-D}\,
     F_{\tau}(tb^{-2}f(b),\epsilon f(b),T f(b)).
\label{eq:4c}
\ee
Here $\Delta N = N-N_{\rm F}$, $b$ is an arbitrary length scaling 
factor, $f(b)=b^D g(\ln b)$, 
and $F_{\sigma}$, $F_N$, and $F_\tau$ are scaling functions. 
The quasiparticle relaxation rate is given as a ratio,
\be
\tau_{\rm QP}^{-1} = \epsilon H''(\epsilon)/H'(\epsilon)\quad,
\label{eq:4d}
\ee
where $H'$ and $H''$ are
the real and imaginary part, respectively, of a causal function 
$H(\zeta=\epsilon + i0)$ of complex frequency $\zeta$ that obeys 
a homogeneity law
\bea
H(t,\zeta,T)&=&{\rm const.}\times g(\ln b)
\nonumber\\
   &&+ F_H(tb^{-2}f(b),\zeta f(b),T f(b))\quad.
\label{eq:4e}
\eea
For completeness and later reference we also note that the specific heat
coefficient, $\gamma = \lim_{T\rightarrow 0} C_V/T$, is given by $H(t,0,T)$
and thus obeys a homogeneity law\cite{us_dirty,us_paper_I,us_paper_II},
\be
\gamma(t,T) = {\rm const.}\times g(\ln b) + F_{\gamma}(tb^{-2}f(b),
              Tf(b)) \ .
\label{eq:4f}
\ee
\eml%
By choosing $b$ appropriately, Eqs.\ (\ref{eqs:1}) and (\ref{eqs:3}) are
recovered from Eqs.\ (\ref{eqs:4}). To obtain Eq.\ (\ref{eq:3c}),
one also needs to use the fact that at the disordered Fermi-liquid fixed
point the relaxation rates are proportional to $\epsilon^{3/2}$\cite{Schmid}.

We will first explain the significance of these results, and then
sketch their origin. 
We first consider the dependence of the conductivity, or the DOS,
on $T$ or $\Omega$ or $\epsilon$. 
In disordered metals away from any
quantum critical point one observes the well-known $T^{1/2}$, 
$\epsilon^{1/2}$, or
$\Omega^{1/2}$ behavior in $3$-$D$\cite{LeeRama}. This is one example
of the weak-localization effects mentioned above, and it is due to the
diffusive nature of the electrons. In a diffusive process, frequency
scales like a wavenumber squared, $\Omega \sim b^{-2}$, and in a quantum
problem, $T\sim\Omega$. Accordingly, the dynamical exponent
$z$, defined by $\Omega \sim b^{-z}$, is $z=2$ in such a system. Near a
quantum critical point, this behavior changes to one characterized
by the dynamical critical exponent characteristic of the phase transition.
For the quantum FM transition in disordered system, 
according to a recent theory the dynamical exponent is $z=D$\cite{us_dirty},
which is reflected in Eqs.\ (\ref{eqs:1}) and ({\ref{eqs:4}),
in contrast to Hertz's theory which yielded $z=4$\cite{Hertz}. Similarly,
the correlation length exponent $\nu$, defined by $t\sim b^{-1/\nu}$, is
$\nu = 1/(D-2)$ according to Ref.\ \onlinecite{us_dirty}, while it is
$\nu = 1/2$ in Hertz's theory. The former 
value is reflected in Eqs.\ (\ref{eqs:3}).
Measuring $\nu$ and $z$ directly at the quantum critical point would be
hard, while an observation of the conductivity or the tunneling DOS
on $T$, $\Omega$ or $t$ should be much easier. Observing the
crossover from the usual $T^{1/2}$ weak-localization nonanalyticity to
the $T^{1/3}$ behavior predicted by the present theory may therefore be
the easiest way to experimentally probe the dynamical critical behavior at the
quantum FM transition. Similarly, the linear dependence of
$\sigma$ and $N$ on $t$ at $T=0$ reflects the value of $\nu$. Also notice that
$d\sigma/dt\propto g(\ln(1/t))\ln t$ at $T=0$, see 
Eq.\ (\ref{eq:3a})\cite{FisherLanger}. 

The behavior of the relaxation times $\tau_{\rm ph}$ and $\tau_{\rm QP}$
is very remarkable from a theoretical point of view. In Landau Fermi-liquid
theory, the relaxation rate is proportional to $\epsilon^2$\cite{AGD}, and
in a disordered Fermi liquid one has $\tau_{\rm ph} \propto \tau_{\rm QP}
\propto \epsilon^{3/2}$\cite{Schmid}. The limiting case between Fermi-liquid
and non-Fermi liquid behavior is given by a `marginal Fermi liquid'\cite{MFL}
with a relaxation rate proportional to $\epsilon$ with logarithmic 
corrections. From Eqs.\ (\ref{eq:1c},\ref{eq:1d}) we see
that itinerant electrons at a quantum FM critical point provide a
realization of a marginal Fermi liquid. The logarithmic $T$-dependence of the
specific heat coefficient is consistent with this. 

We now explain the origin of the scaling behavior discussed above.
Microscopically, 
one should consider a theory of interacting electrons in the presence of
quenched disorder and introduce an order parameter field that couples
to the electron spin density. One then obtains a theory for the 
fermionic degrees of freedom that is missing the spin-density channel
of the electron-electron interaction, coupled to a Landau-Ginzburg-Wilson
(LGW) action for the order parameter field. This program has recently
been carried out\cite{us_paper_I}. While this theory, and its
RG analysis\cite{us_paper_II}, 
are necessary for a complete
justification of our results, most of them can be obtained from simple
scaling arguments in conjunction with low-order perturbation theory.
This we will show here, starting with a scaling analysis.

For scaling purposes, we note that the
observables we are interested in can all be expressed in terms of
fermionic correlation functions. It is therefore reasonable to assume that
the scale dimensions of these observables and the underlying fermionic
fields with respect to the magnetic
fixed point are the same as with respect to
the disordered Fermi-liquid fixed point identified
in Ref.\ \onlinecite{us_fermions}. We will
proceed under this assumption and initially
focus on power laws, ignoring all logarithmic corrections. 
All comparisons of Eqs.\ (\ref{eqs:5}) with Eqs.\ (\ref{eqs:1})-(\ref{eqs:4}) 
are thus to be understood as up to logarithms.

The conductivity is a charge current correlation whose scale
dimension with respect to the quantum magnetic fixed point is
expected to be zero. However, $\sigma$ will depend on the critical
dynamics, since the paramagnon propagator enters the calculation of
$\sigma$ in perturbation theory\cite{perturbation_theory_footnote}.
The corrections to the Boltzmann conductivity will further depend on
the leading irrelevant operator, which we denote by $u$. This is again
related to diffusive electron dynamics, and one therefore expects the
scale dimension of $u$ to be the same as in disordered Fermi-liquid theory,
namely, $[u] = -(D-2)$\cite{us_fermions}.
Standard scaling arguments then suggest a generalized homogeneity law
\bml
\label{eqs:5}
\be
\sigma(t,T,\Omega) = F_{\sigma}(tb^{1/\nu},Tb^z,\Omega b^z,ub^{-(D-2)})\quad.
\label{eq:5a}
\ee
With $z=D$ and $\nu = 1/(D-2)$ from Ref.\ \onlinecite{us_dirty} this
is Eq.\ (\ref{eq:4a}).
By putting $b=T^{-1/3}$, and using that $F_{\sigma}(0,1,0,x)$ is an
analytic function of $x$, we obtain Eq.\ (\ref{eq:1a}).

Similarly, the leading correction $\Delta N$ to the DOS
is given by an integral over a four-fermion 
correlation function whose diffusive dynamics lead to 
$\Delta N \sim b^{-(D-2)}$. This yields Eq.\ (\ref{eq:4b}),
\be
\Delta N(t,\epsilon,T) = b^{-(D-2)}\,F_N(tb^{1/\nu},\epsilon b^z,Tb^z)\quad.
\label{eq:5b}
\ee

Both the dephasing rate and the QP relaxation rate are dimensionally
frequencies. Effectively, their scale dimensions are given by the
critical time scale, $[\tau^{-1}]=D$\cite{us_paper_II}. This implies
Eqs.\ (\ref{eq:4c},\ref{eq:4d},\ref{eq:4e}),
\be
\tau(\epsilon,T) = b^{D}\,\tau(\epsilon b^D,Tb^D)\quad.
\label{eq:5c}
\ee

Finally, the specific heat coefficient is expected to have zero 
scale dimension, as in Ref.\ \cite{us_fermions}. We thus have
Eq.\ (\ref{eq:4f}),
\be
\gamma(t,T) = \gamma(tb^{1/\nu},Tb^z)\quad.
\label{eq:5d}
\ee
\eml%

The logarithmic corrections to scaling in Eqs.\ (\ref{eqs:1})-(\ref{eqs:4}) 
can be understood in terms of Wegner's general classification\cite{Wegner}. 
The first kind is
due to resonance conditions between scale dimensions, which lead to simple
logarithms. The second kind is due to marginal operators, which can lead to
complicated functions of logarithms. Here,
both mechanisms are operative. The scale dimension of the DOS
correction, $[\Delta N] = D-2$, is the same as that of the relevant
operator $t$, $[t] = 1/\nu = D-2$. Further, the correction to the
conductivity is proportional to the irrelevant variable $u$, and
$[tu] = [\sigma] = 0$. These `resonances' lead to the simple logarithms
in Eqs.\ (\ref{eq:4a}) and (\ref{eq:4b}). 

The more complicated logarithms embodied in the function $g(\ln b)$
are due to an effectively marginal operator that is due to the
presence of two frequency scales in the problem, namely the critical one with
$z=D$, and a diffusive one with $z=2$. A full renormalization
group analysis of this problem will be presented elsewhere\cite{us_paper_II},
here we resort to low-order perturbation theory to make the result
plausible. Perturbation theory is also useful for making our general results 
plausible. Consider the DOS as an 
example. To lowest order in both the disorder and the spin-triplet interaction
amplitude $K_t$, the DOS correction in a disordered Fermi
liquid is well known to have the form\cite{LeeRama}
\bml
\label{eqs:6}
\be
\Delta N(\epsilon) \propto G^2\int d{\bf p}\,{\rm Re}
    \int_{i\Omega\rightarrow\epsilon+i0}^{\infty}
   d\omega\ K_t\,\left[{\cal D}({\bf p},\omega)\right]^2\ .
\label{eq:6a}
\ee
Here ${\cal D}$ is the basic diffusion propagator or `diffuson',
\be
{\cal D}({\bf p},\omega) = 1/({\bf p}^2 + GH\vert\omega\vert)\quad,
\label{eq:6b}
\ee
with $G\propto 1/k_{\rm F}\ell$ the disorder parameter, and $H$
proportional to the specific heat coefficient $\gamma$,
see Eqs.\ (\ref{eqs:4}). 
In a theory that keeps $K_t$ to all orders and thus is capable of 
describing magnetism, $K_t$ gets replaced by the paramagnon propagator
${\cal M}$, which has the structure\cite{us_dirty}
\be
{\cal M}({\bf p},\omega) = 1/[t + a_{d-2}\vert{\bf p}\vert^{d-2} + a_2{\bf p}^2 
   + a_{\omega}\vert\omega\vert/{\bf p}^2] \ .
\label{eq:6c}
\ee
The correction to $N$ thus reads
\be
\Delta N(\epsilon) \propto G^2 \int d{\bf p}\,{\rm Re}
  \int_{i\Omega\rightarrow\epsilon+i0}^{\infty} 
  \hskip -20pt 
  d\omega\ 
  \left[{\cal D}({\bf p},\omega)\right]^2\,{\cal M}({\bf p},\omega).
\label{eq:6d}
\ee
The leading correction to $G$ has the same
qualitative behavior as $\Delta N$, Eq.\ (\ref{eq:6d}), while the one to $H$ is
\be
\Delta H(\Omega) \propto G \int d{\bf p}\ \frac{1}{\Omega}\int_0^{\Omega}
   d\omega\ {\cal D}({\bf p},\omega)\,{\cal M}({\bf p},\omega)\ .
\label{eq:6e}
\ee
\eml%
The interesting effects we discuss in this paper result from the
nonanalytic $\vert{\bf p}\vert^{d-2}$ term in Eq.\ (\ref{eq:6c}). We
therefore need to investigate the conditions under which this term
dominates over the analytic ${\bf p}^2$ dependence.
If we measure ${\bf p}$ and $\omega$ in units of $k_{\rm F}$
and $\epsilon_{\rm F}$, respectively, and remember that the
dimensionless spin-triplet interaction is of $O(1)$ near a FM
transition, then the coefficients $a_2$ and $a_{\omega}$ are of $O(1)$,
while $a_{d-2}$ is proportional to $1/k_{\rm F}\ell$. If we scale the
wavenumber with the correlation length $\xi$, this means that the nonanalytic
term will dominate over the analytic one
when $\xi \agt \ell$. With $\nu = 1$ in $D=3$, this translates
into $t\alt 1/k_{\rm F}\ell$. For typical values of the disorder,
$k_{\rm F}\ell \approx 10$, this means that the nonanalytic term will
dominate everywhere in the critical region. For less disordered samples,
our effects will be present in a correspondingly narrower region around
the critical point.

Performing the integral in Eq.\ (\ref{eq:6d}) yields
$\Delta N \propto {\rm const.} + \epsilon^{(D-2)/D}$, while from
Eq.\ (\ref{eq:6e}) we find 
$\Delta H \propto\ln\Omega$. Inserting the latter result
back into the perturbative expressions leads to more complicated logarithms.
Of course, insertions are at best indicative of the behavior at higher
order, a complete treatment needs to take into account vertex corrections
as well. We will report on the details of such a complete analysis
elsewhere\cite{us_paper_II}. The result is that the fully renormalized
$H$ turns into $H\,g(\ln(1/\Omega\tau)$. Similarly, the coefficient
$a_{d-2}$ in the paramagnon propagator gets renormalized to
$a_{d-2}/g(\ln(k_{\rm F}/k))$\cite{us_IFS}. 
Doing the integrals then yields expressions
that are consistent with Eqs.\ (\ref{eqs:1}) and (\ref{eqs:3}). The remarkable
claim that one-loop perturbation theory is exact apart from
logarithmic corrections finds its explanation in the RG analysis of
Refs.\ \onlinecite{us_paper_I,us_paper_II}.

We gratefully acknowledge helpful conversations and correspondence with 
Lior Klein and Achim Rosch. This work was
supported in part by the NSF under grant Nos. DMR--98--70597 and 
DMR--99--75259, by the DFG under grant No. Vo659/2, and by
the EPSRC under grant No. GR/M 04426.
Part of this work was performed at the Aspen Center for Physics.

\vfill\eject

\vskip -5pt
\begin{references}
\b{Fisher} See, e.g., M.E. Fisher in {\it Advanced Course on Critical
 Phenomena}, F.W. Hahne (ed.), Springer (New York 1983).
\b{us_ernst} T.R. Kirkpatrick and D. Belitz, J. Stat. Phys. 
 {\bf 87}, 1307 (1997).
\b{LeeRama} For reviews, see, B.L. Altshuler and A.G. Aronov in 
 {\it Electron-electron interactions in disordered systems}, A.L. Efros and 
 M. Pollak (eds.), North Holland (Amsterdam 1985); P.A. Lee and 
 T.V. Ramakrishnan, Rev. Mod. Phys.  {\bf 57}, 287 (1985).
\b{us_fermions} D. Belitz and T.R. Kirkpatrick, Phys. Rev. B {\bf 56}, 6513
 (1997).
\b{non-FL} See, e.g., the ITP Conference Proceedings published in J. Phys.
 Cond. Matt. {\bf 8} (1996).
\b{us_dirty} T.R. Kirkpatrick and D. Belitz, Phys. Rev. B {\bf 53}, 14364
 (1996).
\b{us_paper_I} D. Belitz, T.R. Kirkpatrick, Maria Teresa Mercaldo, and Sharon
 L. Sessions, cond-mat/0008061
\b{us_paper_II} D. Belitz, T.R. Kirkpatrick, Maria Teresa Mercaldo, and
 Sharon L. Sessions, unpublished results.
\b{Hertz} J.A. Hertz, Phys. Rev. B {\bf 14}, 1165 (1976).
\b{Lonzarich} C. Pfleiderer, G.J. McMullan, S.R. Julian, and G.G. Lonzarich,
 Phys. Rev. B {\bf 55}, 8330 (1997).
\b{CKL} C. Castellani, G. Kotliar, and P.A. Lee, Phys. Rev. Lett. {\bf 59},
 323 (1987).
\b{Schmid} A. Schmid, Z. Phys. B {\bf 271}, 251 (1974).
\b{FisherLanger} Although this is reminiscient of the behavior predicted for
 the thermal ferromagnetic transition by P.G. de Gennes and J. Friedel, J.
 Chem. Phys. Solids {\bf 4}, 71 (1958) (see also M.E. Fisher and J.S. Langer,
 Phys. Rev. Lett. {\bf 20}, 665 (1968)), the underlying physics is very
 different since our effect is entirely dynamical in origin. 
\b{AGD} A.A. Abrikosov, L.P. Gorkov, and I.E. Dzyaloshinski, {\it Methods
 of Quantum Field Theory in Statistical Physics}, Dover (New York 1975).
\b{MFL} C.M. Varma, P.B. Littlewood, S. Schmitt-Rink, E. Abrahams, and
 A.E. Ruckenstein, Phys. Rev. Lett. {\bf 63}, 1996 (1989).
\b{perturbation_theory_footnote} This is because the spin-triplet interaction 
 amplitude that enters perturbation theory
 for $\sigma$ in the theory of disordered Fermi liquids\cite{LeeRama} gets
 replaced by the paramagnon propagator in the case of an itinerant
 ferromagnet.
\b{Wegner} F.~J. Wegner, in {\it Phase Transitions and Critical
 Phenomena}, vol.6, C. Domb and M.~S. Green (eds.)
 (Academic, New York 1976).
\b{us_IFS} Certain logarithmic corrections were missed in
 Ref.\ \onlinecite{us_dirty}. Refs.\ \onlinecite{us_paper_I,us_paper_II}
 show how they arise, and that the exact critical behavior is identical
 with that for an unidentified phase transition discussed before by
 T.R. Kirkpatrick and D. Belitz, Phys. Rev. B {\bf 45}, 3187 (1992),
 identifying the transition treated in that paper as the ferromagnetic
 one.
\end{references}
\end{document}